%% file: ms.tex
\documentclass[12pt,preprint]{aastex}

\def\etal{et~al.\ }

\def\hub{\ifmmode H_\circ\else H$_\circ$\fi}
\def\kms{~km~s$^{-1}$\ }

\shorttitle{H$\delta$ in the Integrated Light of Galaxies}

\begin {document}
\title{H$\delta$ in the Integrated Light of Galaxies: What Are We Actually Measuring?}

\author{L. C. Prochaska} 
\affil{Department of Physics and Astronomy, CB 3255, University of 
   North Carolina, Chapel Hill, NC 27599}
\affil{chaska@physics.unc.edu}
\author{James A. Rose}
\affil{ Department of Physics and Astronomy, CB 3255, University of 
   North Carolina, Chapel Hill, NC 27599}
\affil{jim@physics.unc.edu}
\author{Nelson Caldwell}
\affil{Smithsonian Astrophysical Observatory, 60 Garden Street, Cambridge, MA 02138}
\affil{caldwell@cfa.harvard.edu}
\author{Bruno V. Castilho}
\affil{Laboratorio Nacional de Astrofisica/MCT, CP 21, 37500-000 Itajuba, Brazil}
\affil{bruno@lna.br}
\author{Kristi Concannon}
\affil{Department of Chemistry \& Physics, King's College, Wilkes-Barre, PA 18711}
\affil{kdconcan@kings.edu}
\author{Paul Harding}
\affil{Department of Astronomy, Case Western Reserve University, 10900 Euclid Ave, Cleveland, OH 441066}
\affil{harding@dropbear.case.edu}
\author{Heather Morrison}
\affil{Department of Astronomy, Case Western Reserve University, 10900 Euclid Ave, Cleveland, OH 441066}
\affil{heather@vegemite.case.edu}
\author{Ricardo  P. Schiavon}
\affil{Department of Astronomy, University of Virginia,
P.O. Box 3818, Charlottesville, VA 22903-0818}
\affil{ ripisc@virginia.edu}

\begin{abstract}

We present a cautionary study exploring the reliability of the H$\delta$ line in the integrated 
spectra of galaxies for determining galaxy ages. Our database consists
 of the observed integrated spectra of $\sim$120 early-type galaxies, of 7
metal-rich globular clusters in M31 and the Galactic globular cluster 47 Tuc,
and of the open cluster M67.  We have 
measured H$\delta$ using index definitions designed to assess contamination 
from the CN molecule in and around H$\delta$ by choosing combinations of 
bandpasses that both avoid and include a region of CN molecular lines redward 
of H$\delta$. We find systematic differences in the ages derived from H$\delta$
 measurements among the various definitions when extracting ages from H$\delta$
 in old stellar populations with enhanced CN bands due to non-solar abundance 
ratios. We propose that neighboring CN lines have a strong
 effect on pseudocontinuum and central bandpass levels. For stellar populations which
 have non-solar abundance ratios in C and/or N, population synthesis models that do not account for abundance ratio variations cannot reproduce accurately the CN 4216 \AA \ band, which leads to a corresponding
 inaccuracy in reproducing the various H$\delta$ indices. Hence, caution must 
be used 
when extracting galaxy ages from the H$\delta$ line in old stellar 
populations with significant non-solar abundance ratios. 

\end{abstract}


\section{Introduction}

To derive luminosity weighted mean ages from the integrated light of galaxies, 
the standard method is to
measure the strengths of hydrogen Balmer lines and compare them to stellar 
population synthesis models. Balmer lines are used as age indicators because 
they are sensitive to the temperature of main sequence turn-off stars. 
Balmer lines are most pronounced in A stars and hence peak in the integrated 
light of a stellar population once O and B stars have ended their evolution. 
For ages older than $\sim$ 0.5--1 Gyr, Balmer lines decrease in strength as a stellar population gets older. This is because turnoff stars become dimmer, thus contributing less to the integrated light, and cooler so that Balmer lines become weaker in their spectra.

The redder region of the studied spectrum, which includes the Balmer line H$\beta$, 
has been extensively studied, as the strongest absorption lines are relatively uncrowded compared to regions blueward making them less 
difficult to measure. In particular, the Lick/IDS index system \citep{Fab85, 
Gor93, Wor94} concentrate on
prominent features that lie in this red region \citep{Wor94}. 
Consequently,
H$\beta$ has been widely used as an age indicator. It is only slightly 
contaminated by metallic lines and hence relatively insensitive 
to the age/metallicity degeneracy and to 
abundance ratio affects \citep{Wor94, TB95, Tra00} compared to the more 
crowded Balmer lines in the blue.  These 
characteristics
establish it as well-suited for age diagnostic purposes. In the past two 
decades, 
indices bluer than ~4500 \AA \ have received attention as well, and 
age studies have been expanded to include  higher order Balmer 
lines \citep{Ros85, Ros94, JW95, VA99, Vaz01, CRC03}. Recent studies have 
included newer Lick-style indices defined in \citet{WO97} 
(hereafter WO97) that 
focus on the higher order Balmer lines H$\delta$ and H$\gamma$ \citep{Vaz99,
Sch02}. In addition, \citet{VA99} and \citet{Vaz01} have 
investigated specifically tuned indices for H$\gamma$, and Rose(1985, 1994)
has introduced line ratio indices for comparing Balmer lines with neighboring 
metal lines in galaxies.

Higher order Balmer lines, especially H$\delta$, become necessary for 
studies at significant lookback times, since H$\beta$ is redshifted into an
increasingly problematic night sky spectrum at even modest redshift.
H$\delta$ and other higher order Balmer lines are also important for
breaking the degeneracy 
between age and horizontal branch morphology (and other hot star population) effects \citep{LR03, 
Sch04b}, and to distinguish single age 
populations from multi-population scenarios \citep{Sch06}.  In other
words, to move beyond establishing simply a luminosity-weighted mean age and
metallicity of a galaxy from its integrated spectrum requires the more subtle
undertaking of comparing the strengths of different Balmer lines with 
population models based on different star formation histories.
Additionally, higher order Balmer lines are less affected than H$\beta$ by 
line core fill-in due to emission from
ionized gas (WO97). The primary drawback of using higher order Balmer 
lines is the crowded 
blue region of the spectrum in which they lie, making measurements of these 
lines, uncontaminated by neighboring features, difficult.

Although H$\delta$ appears in some respects to be the cleanest of the higher 
order Balmer lines, it is 
still 
at risk for contamination, especially in older populations where the feature is
 weak. In the pseudocontinuum regions near the H$\delta$ feature 
there are several Fe absorption lines which create sensitivity to metallicity,
 hence increasing the age/metallicity degeneracy. Moreover, several studies 
have mentioned the possibility for nearby CN molecular lines affecting the 
H$\delta$ measurements. WO97 note the overlap of CN 
absorption
 on the H$\delta$ feature, but show no effect on the measured abundances. More 
recently, \citet{Sch02} addressed the influence of CN and CH lines on 
indices in 
the blue and suggested that CN lines could have affected previous age 
estimates for giant elliptical galaxies. \citet{Dre04} also noted a 
depression in the integrated red continuum near H$\delta$ in models 
with increased  CH and CN lines. 

 The possibility of CN molecular lines overlapping with the H$\delta$ passbands
causes concern for studies of intermediate age and old stellar populations,
where 
CN-strong giant branch stars contribute significantly to the integrated light. 
Because most
stellar population synthesis models rely on solar abundance 
ratio isochrones and stellar spectral libraries (see, however, \citet{Kor05})
and massive ellipticals in general exhibit non-solar 
abundance ratios \citep[][and references therein]{Wor92, Kun00}, 
CN molecular lines may not be modelled accurately in these 
stellar populations.
Evidence for the impact of CN lines on blue absorption line indices has recently been provided by \citet{Pro05}, in a study of the Ca4227 index. They found a notable difference in calcium abundance estimates for early-type galaxies when using an 
index definition with only a red continuum passband, Ca4227$_r$, as opposed to 
the conventional Lick style index. This effect is due to a depression of the blue pseudocontinuum of the Ca4227 index, caused by the presence of a CN bandhead in the pseudocontinuum window. This CN bandhead, located at $\lambda$ 4216, is avoided in the Ca4227$_r$ measurement, which makes this newly defined index relatively more immune to the effect of CN contaminination and therefore a cleaner indicator of calcium abundance. Although the CN 
molecular band is most pronounced at the bandhead located at 4216 \AA, it 
extends blueward to the H$\delta$ feature. Thus, a similar effect is possible 
for 
H$\delta$ measurements. 

This paper provides a cautionary study that characterizes the influence of non-solar abundance ratios on 
H$\delta$-derived ages in the integrated light of early-type galaxies. We 
measure H$\delta$ 
using a variety of index measurements designed to assess contamination from CN 
molecular lines in and around the H$\delta$ feature. In \S 2, we introduce the 
data and models. We discuss the region of H$\delta$ and our index 
definitions  in \S 3. In \S 4 we explain the measurements of indices used in 
this paper. In \S 5 we present our results on the reliability of H$\delta$ as an age indicator, and in \S 6 we briefly examine the implications of the results, providing cases where this study is most relevant and pointing toward possible solutions.

\section{Observational Data and Models}

\subsection{Observations}\label{sect:obs}

The observational data used in this paper consist of integrated spectra of 119 early-type 
galaxies, seven metal-rich globular clusters in M31, the open cluster M67, and 
the Galactic globular cluster 47 Tucanae.   The spectra come from the 
following sources:

Spectra for the 119 early-type galaxies, including M32, come from the work of \citet{CRC03}
(hereafter CRC03) and are fully described in that paper.  Here we briefly
summarize the observational setup.  The spectra were obtained with the 1.5m 
Tillinghast Telescope, FAST spectrograph, and a Loral 512 x 2688 pixel CCD 
\citep{Fab98} at the F. L. Whipple Observatory. The spectra have a dispersion 
of 0.75 \AA/pixel and a resolution of 3.1 \AA \ FWHM, and cover the wavelength
region 3500-5500 \AA.  Young galaxies (EWH$\beta$ $>$ 3.0) and high H$\beta$ 
emission  galaxies (H$\beta$ emission correction $>$ 0.3 \AA \ in EW, see \S ~\ref{sect:emission}) were removed from the original CRC03 sample of 175 galaxies.

Spectra for globular clusters in M31 were acquired with the Hectospec multi-fiber spectrograph \citep{Fab05} on the MMT at Mt. Hopkins, AZ. The wavelength range covered is
3700-9150 \AA, at a 
dispersion of 1.2 \AA/pixel and a resolution of 4.5 \AA \ FWHM.  Seven old, 
metal rich star clusters were randomly selected from a large sample of clusters observed for another project.

The integrated spectrum of the open cluster M67 comes from \citet{Sch04a}.  
Briefly, the integrated spectrum of M67 was 
constructed from individual spectra of cluster members at representative
locations in the color-magnitude diagram. The spectra were 
obtained with the same observational setup as for the CRC03 galaxy spectra 
above.  Details of how the individual spectra were weighted and then combined
into a representative integrated spectrum of M67 are given in 
\citet{Sch04a}.

The integrated spectrum of the metal-rich Galactic globular cluster 47 Tuc
comes from the \citet{Sch05} library of integrated spectra of Galactic 
globular clusters.  The integrated spectra were acquired with the R-C
spectrograph on the Blanco telescope at the Cerro Tololo Inter-American
Observatory and a Loral 3Kx1K CCD.  The wavelength coverage is 3360-6430 \AA,
at a dispersion of 1.0 \AA/pixel, and a spectral resolution of $\sim$3.1 \AA.
To obtain a representative integrated spectrum of 47 Tuc, the slit was
trailed over the central core diameter of the cluster, as is further described
in \citet{Sch05}.

\subsection{Population Synthesis Models}\label{sect:model}

To extract the ages, as well as abundances of Fe and other elements, from 
integrated light requires comparison of the observed line strengths in 
integrated spectra to the predictions of a grid of stellar population synthesis
models covering a range in age and chemical composition. In this paper we use 
the 
population synthesis models of \citet{Vaz99} which are fully described in that 
paper \footnote{Vazdekis models are available at http://www.iac.es/galeria/vazdekis/}. 
In this 
paper, we used the Vazdekis model spectral energy distributions (SED's) based on  the stellar library of \citet{Jon98} and computed for a Kroupa Revised power law
IMF with an exponent equal to -1.3, in an age range where SED's are considered reliable. The spectral resolution of Vazdekis model SED's is 1.8 A (FWHM). The result is a 
set of output integrated spectra covering a range in age and metallicity, which
can then be smoothed to match the resolution of the observational data.

\subsection{Synthetic Spectra with CN Abundance Anomalies}\label{sect:synth}

It is well known that massive early-type galaxies are characterized by enhanced
 CN abundances in their spectrum, as compared to stars with solar abundance 
ratios \citep{Tra98, Wor98}. Spectral libraries from which population 
synthesis models are based, such as those mentioned in the previous section, 
lack stars with a CN enhancement similar to that found in early-type galaxies. 
To investigate the behavior of CN molecular lines in the region of H$\delta$, 
it is therefore necessary to turn to spectrum synthesis based on highly 
developed model atmospheres and a complete database of atomic and molecular 
data. 

In the present investigation, we use the synthetic spectra computed by Bruno Castilho \citep[see][for a discussion]{Sch02}. These spectra were computed using the Kurucz model atmospheres and an accurate and complete list of absorption line wavelengths and opacities. The computations were performed at very high resolution (0.04 \AA \ FWHM) and the spectra were later rebinned and gaussian convolved to the dispersion and resolution of the Vazdekis models. The stellar parameters considered were those of typical red giant (T$_{eff}$=4750 K, log$g$=2.5), turnoff (6000 K, 4.0), and subgiant (5250 K, 3.0) stars in the globular cluster 47 Tuc. The spectra were computed for an overall metallicity of
[Fe/H]=-0.5, using two different 
abundance patterns: one for solar chemical composition (scaled to [Fe/H] = -0.5
 dex) and the other for the abundance pattern of CN-strong/CH-weak stars in 
metal-rich Galactic globular clusters, specifically, [C/Fe]=-0.2 and 
[N/Fe]=0.8. In the case of metal-rich Galactic globular clusters, the abundance
 pattern is characterized by a high nitrogen abundance and a mild carbon 
underabundance, rather 
than an overabundance in both C and N, such as might be expected in early-type 
galaxies. The net effect, however, is to enhance the opacity due to the CN 
molecule (since N is fractionally increased at a higher level than C is 
decreased), thus producing the desired result.

\section{Measurements of the H$\delta$ Line Strength}\label{sect:measure}

\subsection{The Spectral Region Around H$\delta$}\label{sect:specreg}

In the most straightforward approach to measuring the strength of absorption lines in 
integrated light, an equivalent width index for a spectral feature (or a 
magnitude equivalent of it) is
constructed from the pseudocontinuum flux (determined from bandpasses to the
blue and red end of the feature of interest) and from the flux in a bandpass
centered on the feature itself.   We refer to the resulting index as a 
pseudo-equivalent width measurement, because only a pseudocontinuum is
measured in the two sidebands; the true continuum is usually impossible to
locate at the spectral resolution typically used in galaxy integrated spectroscopy. For the same reason, it is impossible to avoid blending effects in the index bandpass. Thus, in designing index definitions of this 
style, it is important to isolate, to the extent possible, the relevant feature 
from other nearby features that could overlap in 
the bandpass of the feature itself or contaminate the equally important 
pseudocontinuum bandpasses.

\begin{figure}
\plotone{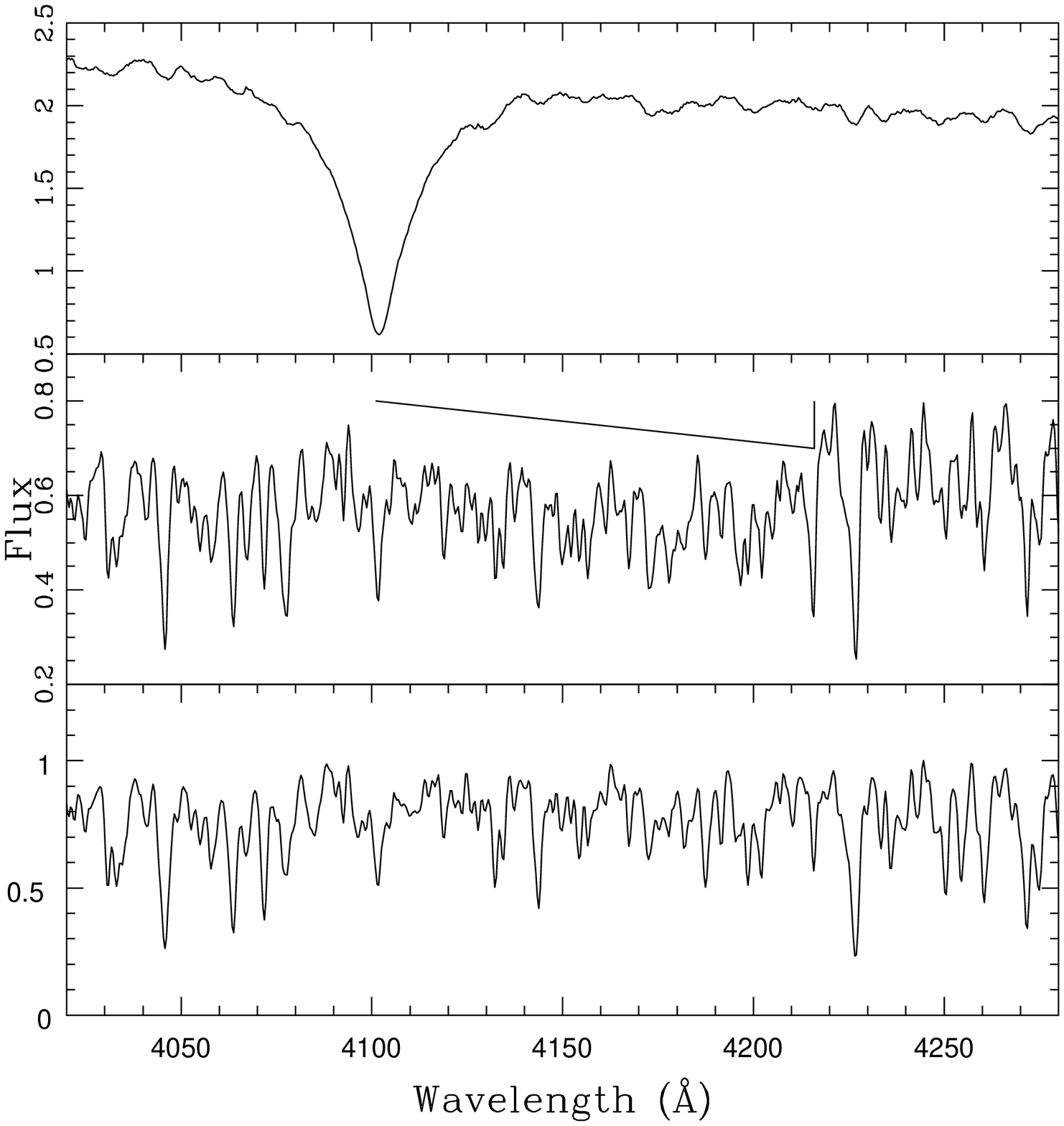}
\caption{Representative spectra of three stars in the region of the H$\delta$ absorption feature illustrate the difficulty in defining an index that cleanly measures the feature. The top spectrum is of the A star H106591, the middle spectrum is of the G giant H210807, and the bottom spectrum is of the K dwarf H10780. The wedge in the middle spectrum indicates the general continuum depression caused by the CN molecular band with bandhead at 4216 \AA. All spectra are from the Indo-US Library of Coude Feed Stellar Spectra and are at a resolution of 1.2 \AA \ FWHM. }
\label{fig:spectra3}
\end{figure}

\begin{figure}
\plotone{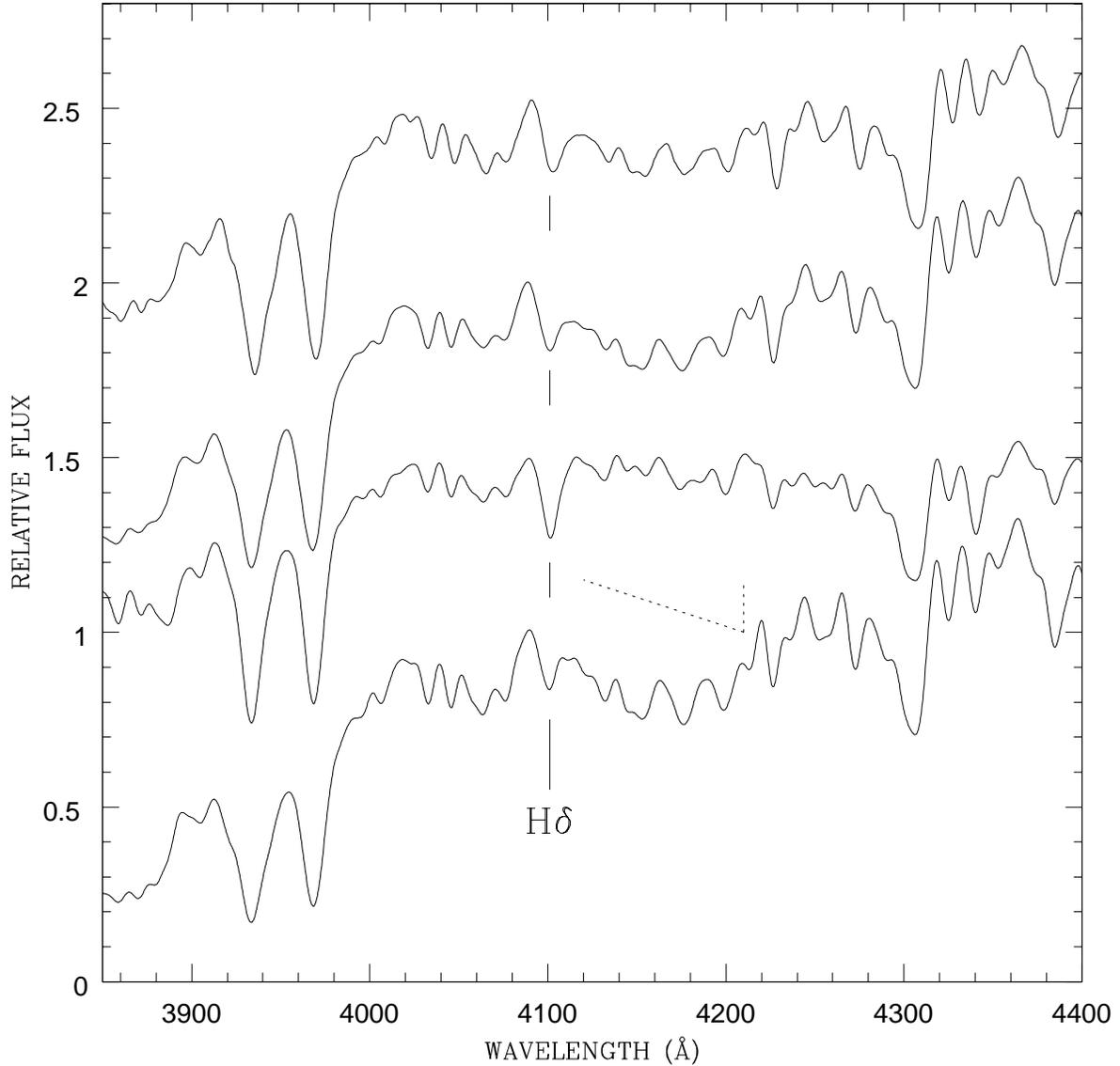}
\caption{Integrated spectra of the elliptical galaxy NGC~821 (top spectrum), 
and the M31 globular cluster 112-174 (second from top) are compared with the 
spectrum of the early G dwarf HD10307 (second from bottom) and that of the
K giant HD4128 (bottom spectrum). Vertical lines denote the H$\delta$ feature.  The depression in the pseudocontinuum 
redward of H$\delta$, indicated by the wedge in the bottom spectrum, is evident in all spectra except for that of the G dwarf. All spectra are at an effective resolution of 3.1 \AA \ FWHM and 230 \kms velocity dispersion.}
\label{fig:Fig2}
\end{figure}

Because it is located in a very crowded spectral region, the H$\delta$ line ($\lambda$4101 \AA) is difficult to isolate from nearby lines, especially in the 
spectra of older populations where H$\delta$ is weaker compared to neighboring
metal lines.
The middle and bottom panels of Fig.~\ref{fig:spectra3}, which show stellar spectra from the 
Indo-US Library of Coude Feed Stellar Spectra \footnote{The Indo-US Library of Coude Feed Stellar Spectra can be found at http://www.noao.edu/cflib/} for a late G giant and a K dwarf 
respectively, illustrate this difficulty. Fe lines are apparent  in the 
spectra, most notably at 4045, 4063, 4072, 4119, 4132, and 4144 \AA \
\citep{Tho04}.  These Fe lines produce absorption in the continuum
sidebands or in the feature bandpass itself, and thereby contribute to a
metallicity sensitivity in the H$\delta$ index.  A result is the well-known
age-metallicity degeneracy effect that complicates stellar population synthesis studies
(e.g. Worthey 1994).  In principle, if the `contamination' of other
features in either the sidebands or central bandpass of an H$\delta$ index
is properly accounted for in stellar population modelling, then reliable ages
and metallicities can indeed be extracted from a comparison of the observed
integrated spectrum of a galaxy with spectral indices generated in
population models.  
However, it has been found that the stellar abundance patterns in early-type 
galaxies frequently differ from the solar neighborhood stars on which the models are 
based. In particular, massive, early-type galaxies are enhanced in $\alpha$ 
elements, such as Mg, as compared to Fe \citep[][and references 
therein]{Wor92, Dav93, Tra00, Kun00}. Studies of globular clusters in the Galaxy indicate that most clusters contain substantial numbers of CN-strong stars and in most cases an enhancement of $\alpha$ elements relative to Fe (or perhaps more correctly stated, 
an underabundance of Fe relative to $\alpha$ elements)\citep{Wor98}. Thus, the integrated spectra of most globular clusters show both enhanced CN and $\alpha$ element features. In this regard, the fact that the CN molecular band at $\lambda$4216 \AA, that 
is prominent in
cool, low gravity stars, is very strong in the integrated spectra of old stellar populations presents a more significant problem than the
contribution from Fe lines.  The CN $\lambda$4216 \AA \ molecular band consists
of a spectral feature that covers more than 100 \AA \ blueward of
the bandhead at 4216 \AA.  The band is indicated by the wedge in the middle
panel of Fig.~\ref{fig:spectra3}.  The most obvious difficulty there
is the extension of the band right up through the redward wing of H$\delta$,
thus making it unavoidable that the redward pseudocontinuum sideband for any
H$\delta$ index will contain a contribution from CN.  This situation is further
illustrated in Fig~\ref{fig:Fig2}, where we plot the spectrum of the old 
metal-rich
elliptical galaxy NGC~821, and compare it to the spectrum of a metal-rich
globular cluster in M31, which we will later show has exceedingly strong CN.
The spectra of a G dwarf and a K giant are also plotted as a reference.  A
striking aspect of the galaxy and M31 globular cluster spectra is that while
there is a prominent pseudocontinuum peak blueward of H$\delta$, no such peak
is seen to the red of H$\delta$ (as was pointed out by Dressler \etal 2004).  The fact that the pseudocontinuum redward
of H$\delta$ is also supressed in the K giant spectrum, which has a prominent
CN $\lambda$4216 band, while there is a well-developed red pseudocontinuum peak
in the G dwarf, which has little contribution from CN, certainly suggests that
the the depression of the redward pseudocontinuum peak in the galaxy spectrum
is primarily the result of the CN band.  We clarify this further in
\S~\ref{sect:synthres}.

As mentioned above, CN-contamination of the pseudocontinuum redward of H$\delta$ does not make the definition of a useful  H$\delta$ index necessarily impossible, since the
effect of CN is in principle accounted for in the population synthesis models.
However, the fact that elliptical galaxies are frequently found to exhibit
non-solar abundance ratios makes accurate modeling quite problematic.  In fact,
creating population synthesis models that include the
accurate representation of non-solar C/Fe and N/Fe ratios proves a difficult 
task. To supplement
stellar libraries with stars of enhanced CN molecular lines, it is necessary 
to use stellar atmosphere models which require enormous complexity to include 
molecular lines. Until such models are widely available, it is
important to be aware of the complicating effect of CN lines on H$\delta$ 
measurements, as we have previously shown in the case of the Ca$\lambda$4227
line \citep{Pro05}, which has CN contamination in the
blue pseudocontinuum bandpass.  Indeed, the
effect of CN on H$\delta$ indices is the main concern of this paper.

Finally,
the top panel in Fig.~\ref{fig:spectra3} shows the spectrum of an A star that 
is characteristic of light 
which dominates the flux in young ($\sim$1 Gyr old) stellar populations. For 
these populations, the effect of metal line contamination is much less severe 
than in 
the older populations we are concerned with in this study.  Thus for the
remainder of this study we primarily concern ourselves with measuring and
modeling H$\delta$ in old stellar populations.

\subsection{H$\delta$ Index Definitions}\label{sect:definitions}

The chosen width for the index passband depends on several considerations. As 
mentioned in the previous section, it is important to have a sufficiently  narrow index 
passband to isolate the feature of interest. For example, a narrow index 
definition can minimize the effect of the metallicity dependence caused by Fe 
line contamination. However, narrow indices are more 
sensitive to velocity broadening effects in which the flux is redistributed 
from the core into the wings. Additionally, as can be seen in the top panel of 
Fig.~\ref{fig:spectra3}, hot (A type)
stars have extensive pressure-broadened wings. A wide index passband is thus important for measuring 
the entire H$\delta$ feature in spectra of young galaxies in which upper main
sequence stars dominate the integrated light. On the other hand, the bottom 
two spectra of Figs.~\ref{fig:spectra3} and \ref{fig:Fig2} are more characteristic of light from older populations. 
Consequently, WO97 defined two Lick-style indices for H$\delta$. 
H$\delta_A$ 
uses wide passbands to include the extended wings of A stars appropriate for
young stellar populations.  The narrower 
passbands of H$\delta_F$ are designed to measure galaxies whose integrated 
light is dominated by later-type stars in older populations. 
For convenience, these definitions are given in Table ~\ref{tab:indexdef}.

To estimate the impact of CN molecular lines on H$\delta$ index measurements, 
we modify the WO97 H$\delta_A$ and H$\delta_F$ indices, defining 2 variations 
for each one.  In the first variation, we define modified H$\delta_A$ and 
H$\delta_F$ indices that only use the blue continuum bandpass for measuring
the continuum flux.  We refer to these indices as H$\delta_{A}b$ and 
H$\delta_{F}b$ ($b$ for blue).  For comparison, we also define 2 indices that 
use only the red continuum passband for the continuum flux (H$\delta_{A}r$ and 
H$\delta_{F}r$).  This approach has been utilized previously for defining 
variations on the Lick Ca4227 index in \citet{Pro05}.
Naturally, we also measure the original H$\delta_A$ and H$\delta_F$ definitions.
The index definitions that only use one passband to define the continuum flux 
in effect assume a flat continuum in the region of H$\delta$. Hence the 
measurement is essentially a mean flux 
ratio between the line and continuum bandpasses but defined in terms of an
equivalent width.  The point of this approach
is to have index definitions that both avoid and include the region of CN 
molecular lines redward of H$\delta$.  We have defined in a similar manner variations of the H$\beta$ index as a test of our method (see \S \ref{sect:hbres}), with H$\beta$$_b$ using only the blue continuum bandpass of H$\beta$, and H$\beta$$_r$ using only the red continuum bandpass.

\begin{figure}
\plotone{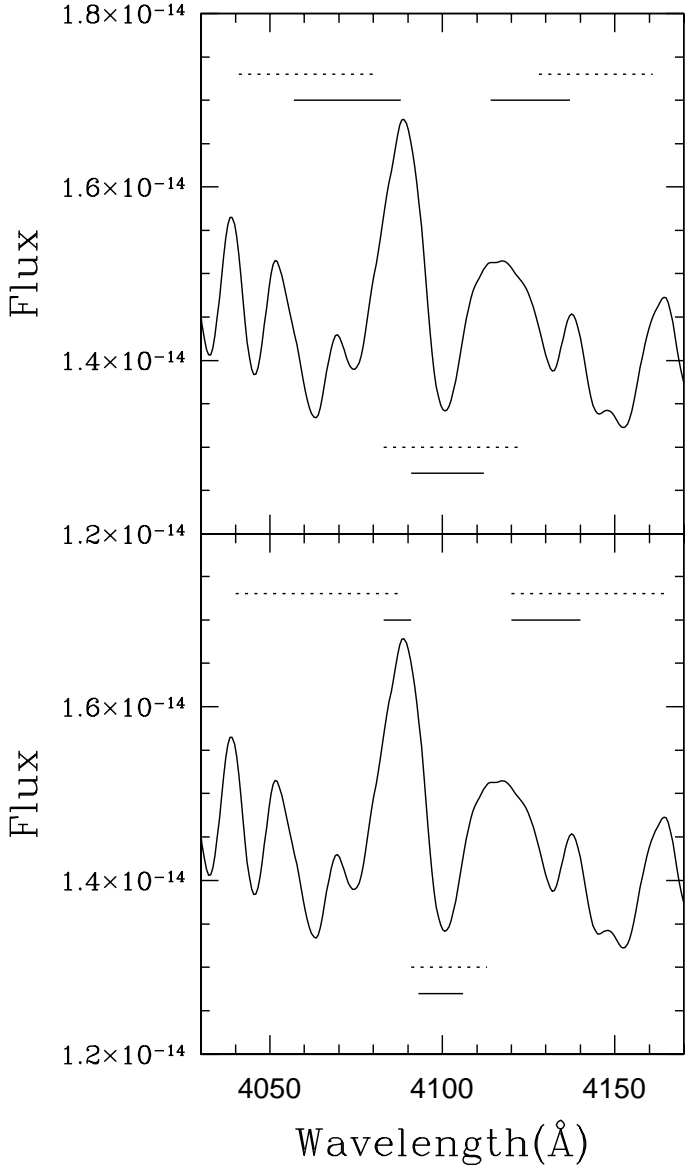}
\caption{The spectrum of the early-type galaxy, NGC 821 is plotted in the vicinity of the H$\delta$ absorption line. Solid lines in the top panel mark the bandpasses used in the narrower WO97 H$\delta$$_F$ index definition,  dashed lines indicate bandpasses of the wider  WO97 H$\delta$$_A$ index definition. The same bandpass placements are used in our index definition variations, H$\delta$$_{A}b$, H$\delta$$_{A}r$,  H$\delta$$_{F}b$, and H$\delta$$_{F}r$ indices. The bottom panel shows bandpasses of the newly defined indices. Dashed lines indicate the wider H$\delta_{Wb}$, H$\delta_{W}$, and H$\delta_{Wr}$ bandpasses and solid lines indicate the narrower  H$\delta_{Nb}$, H$\delta_{N}$, and H$\delta_{Nr}$ bandpasses. The galaxy spectrum has a resolution of 3.1 \AA \ FWHM and 230 \kms velocity dispersion.}
\label{fig:indexdef}
\end{figure}

To find an index definition that best measures the H$\delta$ feature and with the benefit of higher resolution spectra than available for WO97, we have defined a set of new H$\delta$ indices
of varying bandpass widths and placements.
The new indices are defined with the same set of variations as for the WO97
H$\delta_A$ and H$\delta_F$ indices above:
using  blue only,  red only, and both 
continuum bandpasses to define the continuum level.
The new index definitions with the most extreme passband widths (H$\delta$$_Nb$, H$\delta$$_N$, and H$\delta$$_Nr$ for the narrowest bandpasses and H$\delta$$_Wb$, H$\delta$$_W$, and H$\delta$$_Wr$ for the widest), as well as the H$\beta$ and WO97 
H$\delta_A$ and H$\delta_F$ bandpasses, are listed in Table 
\ref{tab:indexdef}. These are also shown in Fig.~\ref{fig:indexdef} (top panel shows WO97 index definitions, bottom panel shows new index definitions), where
index bandpasses (solid lines indicate narrow passbands, dotted lines indicate 
wide passbands) are overplot on the spectrum of an early-type galaxy. These indices are also included in Table \ref{tab:indexdef}.

\section{Spectral Index Measurements and Errors}

All indices used in this study are measured using a modified version of the 
LECTOR program made 
publicly available by Alexander 
Vazdekis\footnote{http://www.iac.es/galeria/vazdekis/models.html}. The measured
 indices for each galaxy in our sample, M67, 47 
Tuc and the seven M31 globular clusters, along with the $\pm$1$\sigma$ 
uncertainties in the indices, are given in Table ~\ref{tab:galaxyindices}. 

The spectra used in this study were taken at a variety of resolutions.  The
lowest resolution spectra are the MMT/Hectospec multi-fiber spectra of star
clusters in M31, taken at a resolution of 4.5 \AA \ FWHM. On the other hand, the 
CRC03
early-type galaxy spectra, although taken at 3.1 \AA \ FWHM resolution, have 
all been smoothed to a common velocity 
broadening of $\sigma$=230 \kms, which at the 4101 \AA \ wavelength of 
H$\delta$
corresponds to a resolution of 8.0 \AA \ FWHM. Whenever spectral indices were 
derived from the data, all galaxy and cluster data were first smoothed to a 
common spectral resolution as specified by the Lick IDS resolution specific for each index \citep{WO97}.

\subsection{Lick Indices}

In this paper we use several key spectral line indices (H$\beta$, Fe4383, and
 Mg~b) that are defined in 
the Lick system by \citet{Wor94} and in WO97 (H$\delta_A$ and H$\delta_F$) in addition 
to the newly defined indices for H$\delta$ and H$\beta$.  It is important to 
note that the Lick style indices used here are not entirely on the standard Lick
system. The true Lick system originated with the IDS detector,  in which the 
spectral resolution varies with wavelength (as described in WO97) and the 
spectra are on the instrumental response of the IDS. The data used in this 
study are flux calibrated and originally acquired at a variety of spectral
resolutions.
To compare the data and models, it is necessary to smooth them to a common 
resolution. As noted in the previous section, we have performed 
smoothing on the data specific to each spectral feature being 
measured in order to place  the data closer to the Lick system. For example, before measuring Fe4383, all data and models were 
gaussian smoothed to a resolution of 9.3 \AA \ FWHM, the Lick resolution at 
4383 \AA.  By working at the specified Lick
resolution for each feature, we bring our data substantially closer to the
original Lick system.  However, we do not adjust our flux-calibrated spectra
and models to the Lick/IDS spectral response.

\subsection{Emission Line Fill-In}\label{sect:emission}

Balmer emission, due to HII regions (as well as diffuse HII in the galaxy), active galactic nuclei, and/or planetary
nebulae, fill in the underlying Balmer absorption 
lines,  causing spuriously older age estimates from the measured spectra 
\citep{Tra00, Dre04}. The H$\delta$ index 
measurements in this study
have not been corrected for emission. However, our 
present study is aimed at looking for a difference in ages for different
H$\delta$ line index definitions, and the presence of emission in H$\delta$ is
unlikely to influence {\it differential} ages.
Regardless, we expect only a small amount of emission 
line fill in at H$\delta$, since the CRC03 sample was itself restricted to
galaxies which have only weak emission, and since
H$\delta$ is less afflicted than lower order Balmer lines due to the steep
Balmer decrement in emission \citep{Ost89, WO97}.
To ensure that emission plays no role in our results, we have 
removed from the original CRC03 sample of 175 galaxies those 
with H$\beta$ emission correction (from CRC03) in excess of 0.3 \AA \ in
equivalent width.  The determination of H$\beta$ emission is described in
CRC03.

\subsection{Continuum Slope Effect}\label{sect:contslope}

Our study examines differences in derived ages among various index 
definitions, some of which are formulated using only one continuum bandpass. While \citet{Vaz99} showed that B-V colors measured in the Jones spectra did not exactly match the known B-V colors of the Jones stars, he suggests that the flux calibration quality is acceptable for spectral regions smaller than \~ 300 \AA. Therefore, this effect has little influence on the present study as the pseudocontinuum and index passbands are relatively close to one another. Nevertheless,
it is important to consider that blue continuum indices may 
give different derived ages than those using the red continuum indices
if there are systematic differences in continuum 
slopes between the data and models. A difference in slope could arise from flux
calibration errors and reddening effects, as well as deficiencies in the models. 

To estimate and correct for the effect of differing continuum slopes on 
H$\delta$ measurements, we have adjusted the slope of the observed spectra to 
have the 
same continuum slope as that of the appropriate model. Observed spectra were 
matched to 
model spectra with similar H$\beta$ and Fe4383 measurements. The ratio of each 
observed spectrum and corresponding model was fit to a line using the 
{\it curfit} 
routine in IRAF, then divided into the original observed spectrum to yield an 
observed spectrum with the same continuum slope as the model spectrum. Because 
ages are derived by overplotting data onto model grid lines, any difference in 
H$\delta$ measurements of various definitions caused by the continuum slope 
will match the difference of the models. As a result, there should be no 
systematic effects on derived ages due to slope effects. All indices in this 
study 
were measured on observed spectra with adjusted slope. We further
explore the possibility of continuum slope errors on the spectral indices in
\S5.3.

\subsection{Errors in Spectral Indices}

Uncertainties in the spectral indices used here are in principle already 
determined by CRC03 for the early-type galaxy sample. However, since we have 
recalculated the Lick Fe4383 and WO97 H$\delta_A$ and H$\delta_F$ indices after
 smoothing to the Lick resolution specified in WO97 and 
have calculated new versions of H$\delta$ and H$\beta$ indices, we determined 
the
 new 
1$\sigma$ uncertainties by calculating the r.m.s scatters in the Lick indices 
in multiple exposures for a subsample of the CRC03 galaxies. From this 
subsample we determined an overall correction factor from the published CRC03
index uncertainties to those in our index definitions carried out at the Lick 
resolution. This approach was carried out for each index definition used in our
 study. In the case of 47 Tuc, M67,
and the M31 globular clusters, where multiple exposures were not available, we 
calculated the uncertainties in the Lick indices from the signal to noise ratio
 per pixel integrated over the specified bandpass (or used, in the case of M67, uncertainties derived in \citet{Sch04a}) and applied the above 
correction factor to other index definitions.

\section{Results}

\subsection{Ages Derived from Worthey \& Ottaviani H$\delta$ Indices}\label{sect:wores}

The primary goal of this paper is to evaluate the reliability of using 
H$\delta$ for determining galaxy ages.  In principle, if H$\delta$ based ages are reliable, then the derived ages should not be
highly sensitive to the exact definition of the H$\delta$ line index.  On the
other hand, if the derived ages change substantially depending upon the exact
bandpasses used in the index, then it is likely that one or more bandpasses
contain a feature that is not being properly modelled, perhaps due to the
presence of non-solar abundance ratios in the galaxy spectra.  As mentioned
before, we are especially concerned about the effect of the CN $\lambda$4216 molecular
band on the red continuum side of H$\delta$. Thus we have defined modifications of
the WO97 H$\delta_A$ and H$\delta_F$ indices which utilize the continuum on
only the blue or the red side of H$\delta$.

\begin{figure}
\plotone{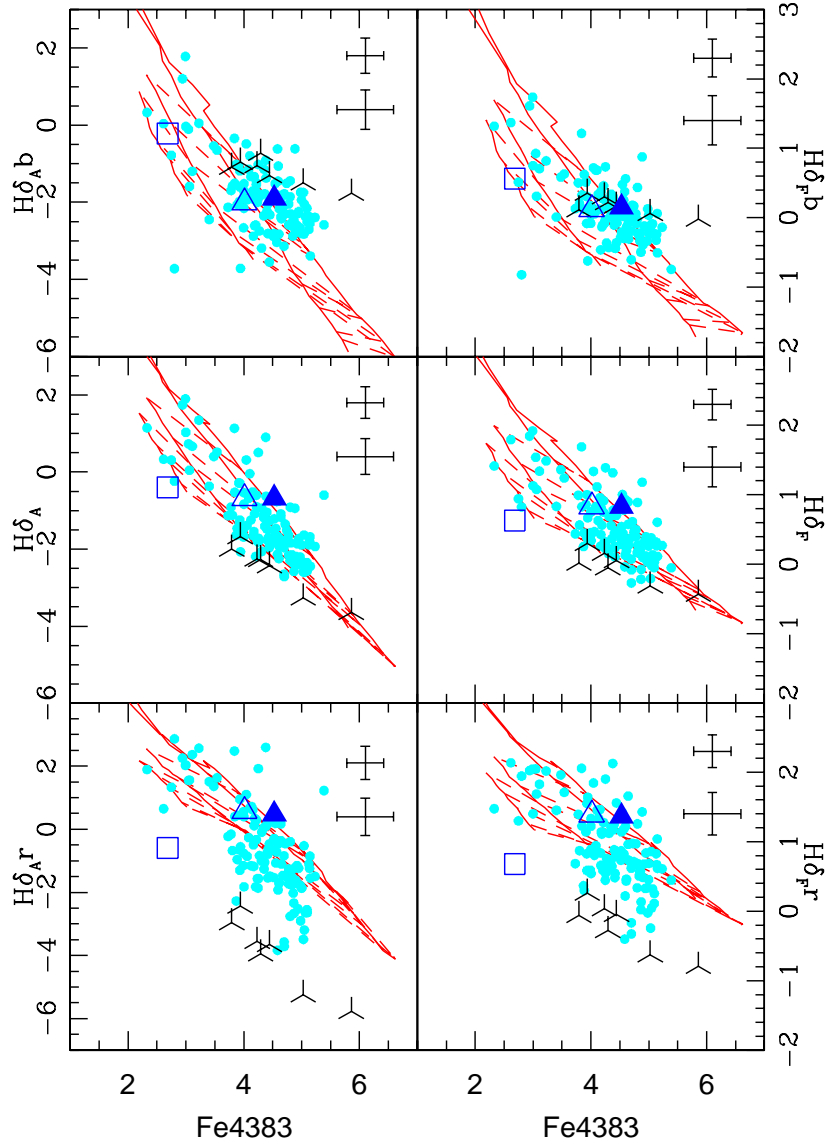}
\caption{H$\delta$ indices are plotted versus Fe4383 for the CRC03 integrated spectra of early-type 
galaxies (small filled circles). Also plotted is a data point for the globular 
cluster 47 Tuc (open square), the open cluster M67 (unfilled triangle), the M31 
globular clusters (skeletal triangles) and M32 (filled triangle).  Overplotted 
on the data points are the model grid lines of constant age and [Fe/H] from 
Vazdekis (1999). The specific metallicities (solid lines) are, from left to 
right: [Fe/H]= -0.68, -0.38, 0.0, +0.2. The ages (dashed lines) are from top to 
bottom:
1.00, 2.51, 3.98, 5.62, 8.91, 11.22, 14.12, 15.85 Gyr. The top panels show the H$\delta$$_A$ (left panels) and H$\delta$$_F$ (right panels) which are measured using only the blue continuum bandpass, while the middle panels show the original WO97 H$\delta$$_A$ and H$\delta$$_F$ index measurements. The bottom panels plot data for the H$\delta$$_A$r and H$\delta$$_F$r indices, which measure H$\delta$ using only the red continuum bandpass. Error bars represent the $\pm$1$\sigma$ errors for the galaxy observations (upper error bars) and M31 clusters (lower error bars).}
\label{fig:wovsfe}
\end{figure}

A common procedure to break the age-metallicity degeneracy in integrated SED's of old stellar populations is to  plot a primarily age sensitive index against a primarily metallicity
sensitive index. Model grid lines are then overlaid to determine the age and
metallicity of the population.  We follow this procedure with the WO97 
H$\delta_A$ and H$\delta_F$ indices, which are plotted versus the metallicity
sensitive Fe4383 index in Fig.~\ref{fig:wovsfe}.  The original WO97 indices,
which utilize both blue and red pseudocontinuum bandpasses, are plotted in
the middle panels, with H$\delta_A$ on the left and H$\delta_F$ on the right.
The CRC03 early-type galaxy data, as well as additional points for the M31
globular clusters, the integrated spectrum of M67, and that of 47 Tuc and of
M32, are all plotted along with the model grid lines.  The constant age and
constant metallicity grid lines do show a useful degree of separation.  However,
they are far from orthogonal, which is presumably due mostly to the large amount
of metal-line contamination in both the H$\delta$ central and continuum bandpasses. In the top and bottom panels of Fig.~\ref{fig:wovsfe} we plot the newly defined
H$\delta_{A}b$ and H$\delta_{F}b$ (top left and right panels) and H$\delta_{A}r$
and H$\delta_{F}r$ (bottom left and right panels) indices versus Fe4383.  To
recap, the top panel indices use only the blue continuum bandpass, while the
bottom panels use only the red continuum bandpass.  If [C/Fe] and [N/Fe]
abundance ratios in the CRC03 galaxies are basically solar, then we would
expect the CN bands to be properly accounted for in the population synthesis
models, in which case H$\delta$ derived ages should be the same in all panels. Instead, there is a large offset in
derived ages, with the H$\delta_{A}r$ and H$\delta_{F}r$ ages being systematically
older than those from the standard WO97 H$\delta_A$ and H$\delta_F$ ages.  In
fact, for the H$\delta_{A}r$ and H$\delta_{F}r$ indices, most of the galaxies 
lie 
completely outside of any reasonable age expectation for these galaxies.  Note
that the metal-rich M31 globular clusters show the effect at an even more
extreme level than for the CRC03 galaxies.  The Galactic globular cluster
47 Tuc is also significantly impacted.  On the other hand, the open
cluster M67 shows no age discrepancy between the two indices.  As well, the
low-mass elliptical galaxy M32 shows only a modest discrepancy in age between 
the two indices. A similar figure substituting Fe5270 for Fe4383 gives an equivalent result.

A natural explanation for the age discrepancies that are found between the
two H$\delta$ index measurements is that the red pseudocontinuum bandpass is
affected by the strength of the CN molecular feature and that [CN/Fe] is
enhanced in the bulk of the CRC03 galaxies and in the M31 globular clusters,
such that the Vazdekis models, based on solar abundance ratios, do not
adequately model the depression in the red pseudocontinuum bandpass that
appears in the observed spectra.  This conjecture is in accord with the finding
that more reasonable ages are derived when the blue pseudocontinuum
bandpass is included (with the WO97 form of the indices). There is compelling
evidence that carbon and nitrogen are, in fact, overabundant in early-type 
galaxies \citep{Tra98, Wor98, Sch06} and in M31 globular clusters \citep{Bur84, Tri89,
Bea04}, although \citet{San03} have found that those effects are less pronounced in cluster ellipticals. In fact, dramatic overabundances in N have been found in M31 globular clusters, while C appears to be normal \citep{Pon98, Bur04}.   Further support for CN as the culprit in the age 
discrepancies between indices lies in the fact that the open cluster M67 is
known from spectroscopy of its giants to have essentially normal C and N 
abundances, while integrated light studies of M32 indicate only modest
departures from mean solar abundances \citep{Wor04, Sch04a, Ros05}, with [N/Fe] essentially solar in M32 \citep{Bou88}.

If we now assume that all the damage caused by CN overabundances is contributed
in the red pseudocontinuum bandpass of the WO97 H$\delta$ indices, it is then
to be expected that both younger and more reliable ages would be obtained if
only the blue pseudocontinuum bandpass is used.  However, while inspection of 
the top panels in Fig.~\ref{fig:wovsfe} does reveal that younger ages are
obtained from the blue-bandpass-only definition for H$\delta$, 
the ages are unreasonably young, this time falling outside the model lines to
the unreasonably young side.  Interestingly, the most reliable ages appear to
be obtained when {\it both} blue and red pseudocontinuum bandpasses are used,
as in the original WO97 definitions.

\subsection{Ages Derived From H$\beta$ Indices}\label{sect:hbres}

\begin{figure}
\plotone{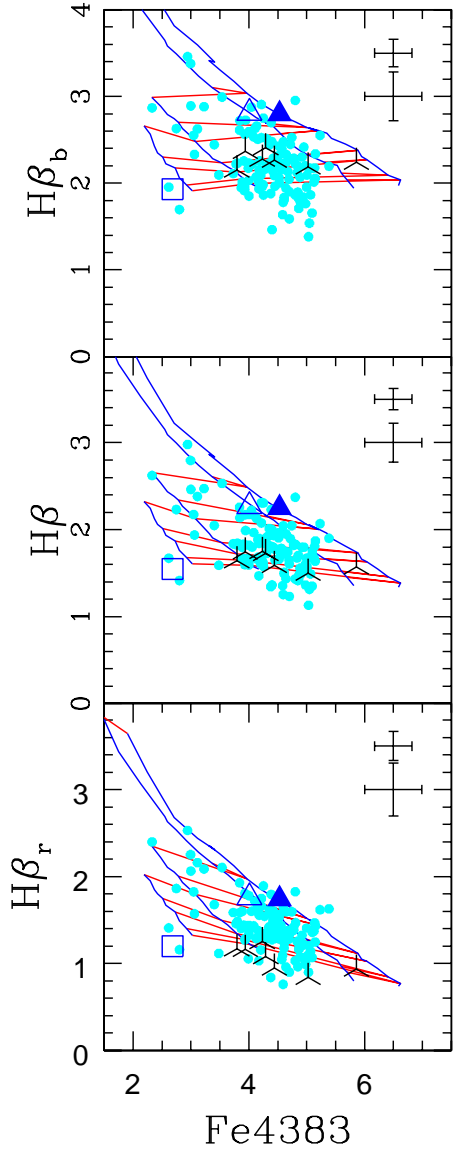}
\caption{H$\beta$ is plotted versus Fe4383 for the same CRC03 galaxies, M31 globular 
clusters, 47 Tuc, M32, and M67 data as in Fig.~\ref{fig:wovsfe}. The top (bottom) panels show H$\beta$ 
measured using only the blue (red) continuum bandpass, while the middle panels
 show the original H$\beta$ measurements.  The model grid lines indicate the 
same ages and 
metallicities as in Fig~\ref{fig:wovsfe}. All symbols and error bars are the same as for Fig~\ref{fig:wovsfe}. }
\label{fig:hbeta}
\end{figure}

Before further investigating why both blue-only and red-only continuum bandpass
definitions of H$\delta$ produce such unreasonable age results, we first 
demonstrate
that the same effect is {\it not} apparent in the Lick H$\beta$ index. Since neither the pseudocontinuum nor the central bandpasses of the H$\beta$ index are contaminated by any very strong lines, the
expectation is that ages derived from H$\beta$ should be independent of
whether a blue only or red only pseudocontinuum bandpass, or both bandpasses,
are used in the index definition.  As seen in Fig.~\ref{fig:hbeta}, indeed 
H$\beta$-extracted ages are largely independent of which pseudocontinuum 
bandpass is used.  There is some tendency for the M31 globular clusters to
exhibit systematically older ages with the red bandpass only (bottom panel
of Fig.~\ref{fig:hbeta}), and younger with the blue bandpass only (top panel), 
but the effect is far smaller than with H$\delta$.
Otherwise, H$\beta$ ages appear robust, regardless of pseudocontinuum bandpass
selection.  Thus the large age variations seen in the different H$\delta$ 
indices is specific to H$\delta$ itself, and not simply an artifact of our 
single continuum bandpass approach.

\subsection{Continuum Slope Revisited}

\begin{figure}
\plotone{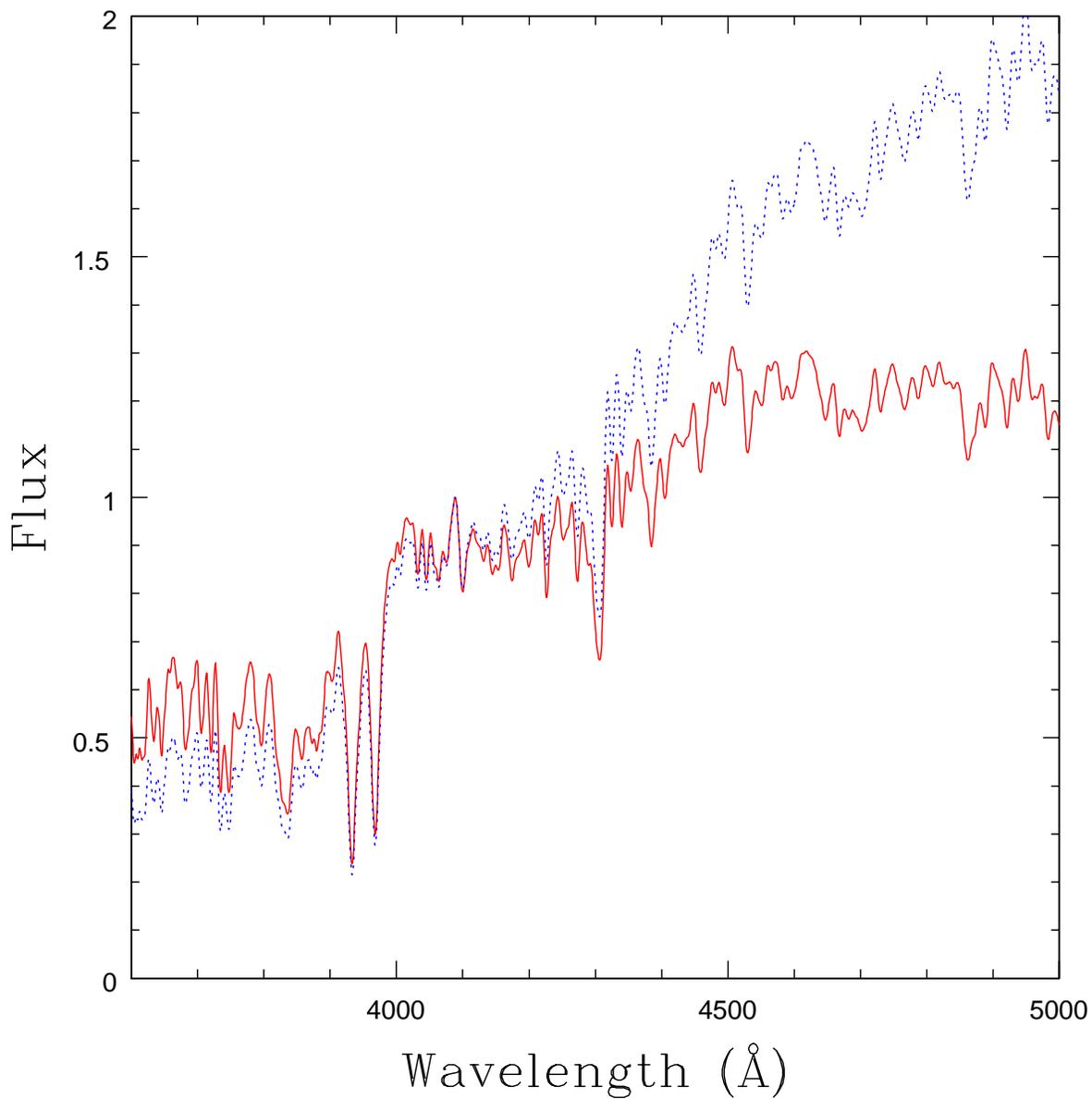}
\caption{Spectrum of early type galaxy, N774, adjusted to have a B-V color
increased by 0.37 mag (blue, dotted) is overlaid on the original spectrum (red, solid) in the
vicinity of the H$\delta$ spectral feature.  
Both spectra
have been normalized to unit flux at 4088 \AA.  The spectra are at a resolution of 3.1 \AA \ FWHM and 230 /kms.}
\label{fig:speccompn774}
\end{figure}

Although we have previously adjusted the continuum slope of our data, as 
discussed in \S~\ref{sect:contslope}, to
match that of the model spectra which they are compared to, it is important to
be confident that the age discrepancies between different continuum
bandpass definitions seen in Fig.~\ref{fig:wovsfe} are not 
caused by an effect that an error in 
continuum slope could have on our single continuum bandpass approach. Therefore, we 
perform a final test by adjusting the slope of our original early-type galaxy data in increments 
until the single-continuum-bandpass index measurements (H$\delta_{A}b$, 
H$\delta_{A}r$, H$\delta_{F}b$ and H$\delta_{F}r$) give a consistent age 
with that of the original WO97 index measurements. The original WO97 indices 
and their variations are measured on the slope adjusted data at each increment 
and then plotted onto a figure similar to Fig.~\ref{fig:wovsfe}. From this, we determine that the
version of the data with a $\sim$40\% increase in slope over the 4400-5500 
\AA \ baseline, i.e., a 0.37 mag change in B-V color, gives a consistent age for
 all indices, which in effect, represents the degree of continuum slope error 
necessary to produce the result we find in Fig.~\ref{fig:wovsfe} on slope error alone.  We show in Fig.~\ref{fig:speccompn774} an  early type galaxy spectrum adjusted to 
have a 0.37 mag increase in B-V color overlaid on the original 
spectrum, both normalized to unity just blueward of H$\delta$. It is clear that 
such a large shift
in continuum slope due to spectrophotometric errors is unlikely. Furthermore, we removed that possibility when we flattened our galaxy spectrum continua to match those of the model spectra before performing our analysis. It is only when the galaxy continuum is 40\% different from the model continuum that the problem arises.

\subsection{Synthesis of the CN $\lambda$4216 Molecular Band Using Spectrum Synthesis From Model Atmospheres}\label{sect:synthres}

We now return to the puzzling result that the most reasonable H$\delta$ ages 
result from using {\it both} pseudocontinuum bandpasses, when in fact our
expectation is that only the red bandpass is profoundly affected by CN 
abundance anomalies.  The fact that blue bandpass only H$\delta$ definitions yield
ages that are substantially too young can result from CN contamination of the
central H$\delta$ bandpass.  Unfortunately, to test this hypothesis directly
using observed stellar spectra from the Solar Neighborhood is not possible, 
since the solar neighborhood is populated by stars with normal CN abundance
ratios.  As a result, we have resorted to investigating the effects of CN on
H$\delta$ indices by computing synthetic spectra in the wavelength region of
H$\delta$, based on model atmospheres and a comprehensive database of atomic
and molecular data.  We have described the synthetic spectra in \S~\ref{sect:synth}.

Computing reliable synthetic spectra from model atmospheres 
of cool stars is a very daunting task, as accurate determinations of thousands of oscillator 
strengths are required to determine the strengths of individual lines. . This difficulty is further exacerbated in the blue, where atomic 
lines are significantly crowded. An additional difficulty associated with computing synthetic spectra of early-type galaxies is knowing the correct abundances of carbon and nitrigen. For early-type galaxies, the individual carbon and nitrogen abundances generally are poorly constrained. Consequently, the correct mixture  of carbon and nitrogen to input to synthesis models for early-type galaxies is not known. However, in this study we are concerned with 
differential effects, namely, the influence of changes in CN-band strengths on the 
H$\delta$ feature, which is possible to investigate using spectrum synthesis from model atmospheres.

\begin{figure}
\plotone{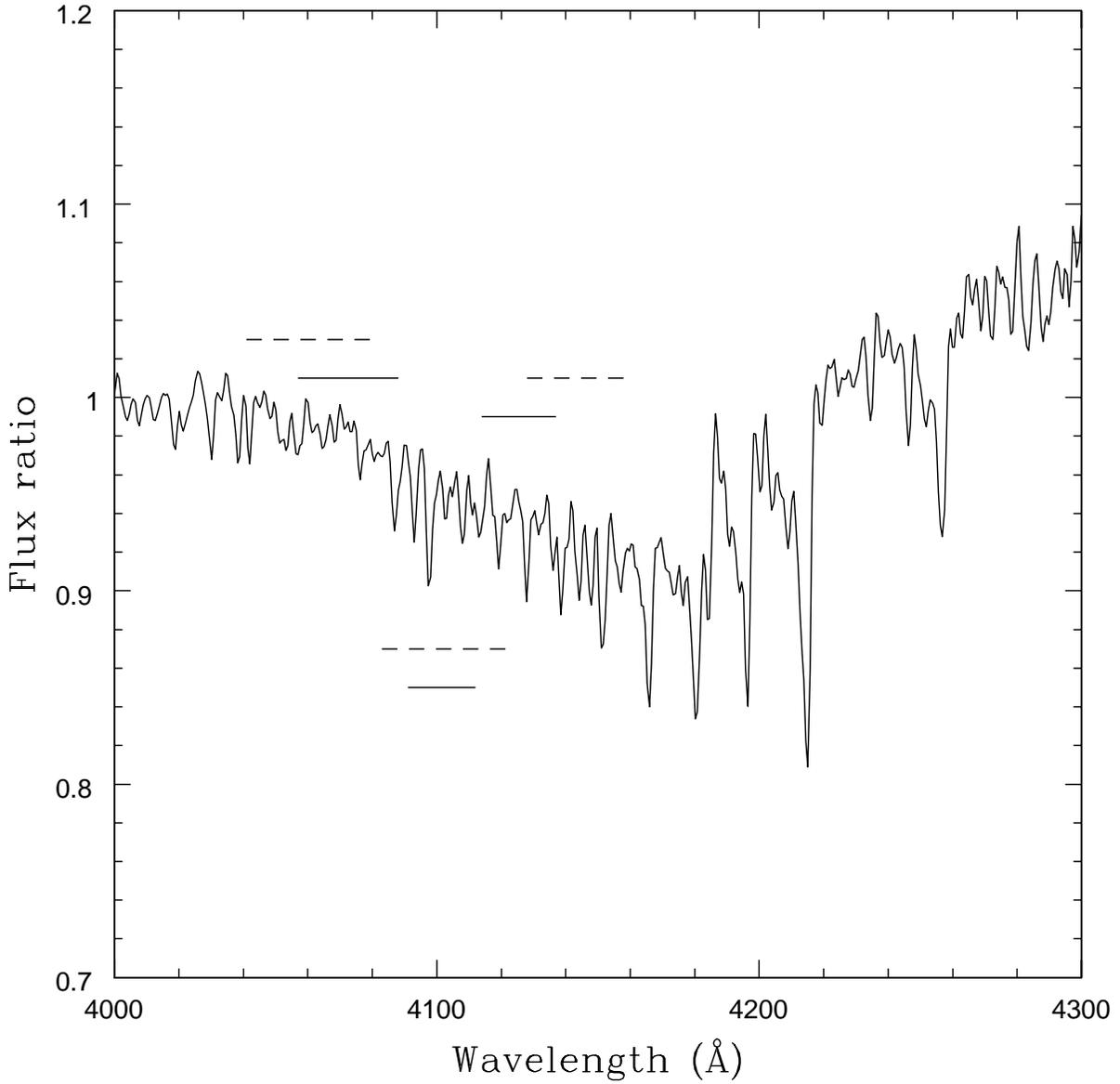}
\caption{The ratio spectrum is plotted for a CN-strong synthetic spectrum divided by a CN-normal synthetic spectrum in the vicinity of H$\delta$. The WO97 H$\delta$ wide and narrow index bandpasses are noted with the same solid and dashed lines as in Fig.~\ref{fig:indexdef}.  The ratio spectrum is plotted at a 
resolution of 1.8 \AA \ FWHM.}
\label{fig:indexdefcn}
\end{figure}

To briefly recap \S~\ref{sect:synth}, synthetic spectra were calculated using 
two different 
abundance patterns, one with solar chemical composition (scaled to 
[Fe/H] = -0.5 dex) and one with the abundance of CN strong/CH weak stars in
metal-rich Galactic globular clusters ([C/Fe] = -0.2 and [N/Fe] = 0.8).  Note
that these abundance ratios were originally chosen to mimic the abundances of 47 Tuc, as determined by \citet{Can98}.  However, the larger N enhancement ratio than C depletion factor will
lead to CN enhancement, which is the desired effect.  
In Fig.~\ref{fig:indexdefcn} we plot the ratio of the two synthetic spectra 
such that the resultant spectrum essentially isolates the effect of CN 
abundance
enhancement on the spectrum of a giant branch star.

The most notable effect in Fig.~\ref{fig:indexdefcn} is indeed a pronounced
depression in the continuum just blueward of the 4216 \AA \ CN bandhead that was previously illustrated by the wedge in Figs.\ref{fig:spectra3} and \ref{fig:Fig2}.  The
depression is highly structured, due to individual rotational transitions,
as well as showing an overall trend of weakening towards the blue.  
While the overall continuum depression has clearly decreased at the wavelength
of H$\delta$, when compared to near the molecular bandhead, there is still a
significant effect at H$\delta$ itself.  More importantly, there is a 
noticeable line just blueward of H$\delta$, which falls within the WO97
H$\delta_A$ and H$\delta_F$ central bandpasses.  To better see the effect, the
WO97 central bandpass and blue and red pseudocontinuum bandpasses have been
marked on Fig.~\ref{fig:indexdefcn} for both H$\delta_A$ and H$\delta_F$
indices.  In fact, the central H$\delta$ bandpass contains two significant
CN lines in the case of H$\delta_F$, and several lines in the case of 
H$\delta_A$.  This fact represents a key setback for finding a clean definition
of H$\delta$, since a substantial CN molecular line falls within 5 \AA \ of
H$\delta$ itself.  With the typical velocity broadening of early-type galaxies,
the CN line will blend with H$\delta$, making the contamination
unavoidable.

To summarize what is evident in Fig.~\ref{fig:indexdefcn}, CN most dramatically
affects the red pseudontinuum bandpasses of the WO97 H$\delta$ indices, 
partially contaminates the central bandpasses, and has virtually no effect on
the blue pseudocontinuum bandpasses.  Accordingly, the strange behavior of
H$\delta$ derived ages in Fig~\ref{fig:wovsfe} can be qualitatively explained 
by non-solar CN abundance ratios in the CRC03 early-type galaxies and in the 
M31 globular clusters.  Specifically, one would expect artificially old ages
to be derived from the red continuum bandpass only H$\delta$ definitions, since
the red bandpass is more heavily afflicted with CN than the central bandpass.
On the other hand, H$\delta$ ages derived from the blue continuum bandpass 
only should
produce spuriously young ages, since the central bandpass itself does suffer
significant CN contamination, while the blue bandpass is relatively clean.
Interestingly, the original WO97 index definitions will produce the
most reasonable ages, since the low CN contamination of the blue 
pseudocontinuum bandpass partially offsets the high contamination of the red
bandpass.  The resultant average roughly matches the CN depression in the
central bandpass, yielding the most reliable ages.  Nevertheless, the exact
degree of cancellation is uncertain and may vary for different population ages, thus causing age determinations from H$\delta$ to be uncertain.

\subsection{Comparison Between Observed and Model Galaxy Spectra}

\begin{figure}
\plotone{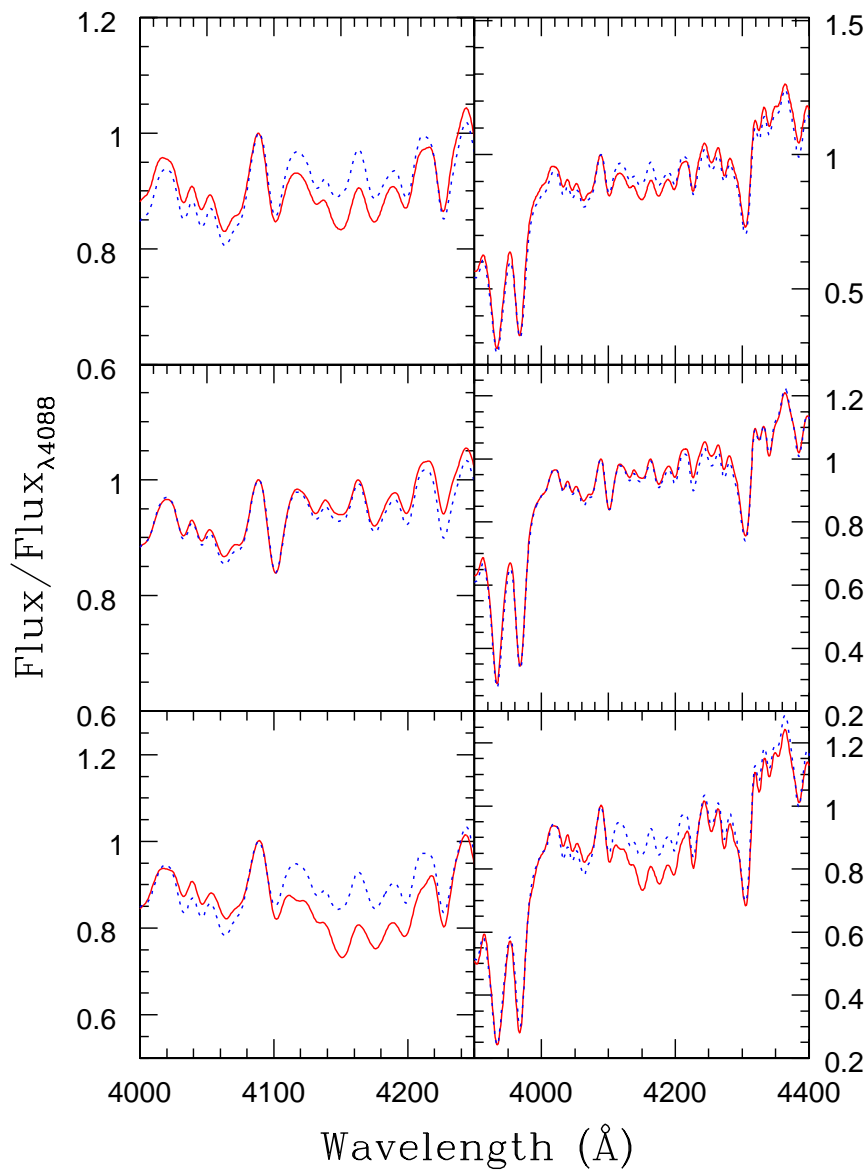}
\caption{Model spectra (blue, dotted) are overlaid on observed spectra (red, solid) in the
vicinity of the H$\delta$ spectral feature.  The left panels show a magnified
view of the H$\delta$ region, while the right panels plot a larger spectral
region.
Top:  Early-type galaxy (NGC 821) spectrum is plotted with a solar 
metallicity, 10 Gyr model spectrum. Middle: Open cluster M67 spectrum is plotted with 
a solar metallicity, 3.98 Gyr model spectrum. Bottom: M31 globular cluster 
(189-240) spectrum is plotted with a [Fe/H]= +0.2, 11.22 Gyr model spectrum. All spectra
have been normalized to unit flux at 4088 \AA. The data have been adjusted 
to share a similar continuum slope as the model spectra of comparible age and 
metallicity. All spectra are at a resolution of 10.9 \AA \ FWHM.}
\label{fig:speccompalldoub}
\end{figure}

In the above section we saw that CN molecular lines prominent in giant stars 
indeed fall into bandpasses 
of H$\delta$ indices, which could in principle compromise H$\delta$ indices if 
the CN is not correctly modeled. But does the integrated light of early-type 
galaxies actually have strong enough CN 4216 bands to affect flux levels in the
 red 
and central bandpasses? To evaluate this, we overplot a \citet{Vaz99} model 
spectrum (as used in Fig.~\ref{fig:wovsfe} and ~\ref{fig:hbeta}) against an 
early-type galaxy spectrum of the same age and metallicity (determined from 
Fe4383 and H$\beta$ index measurements and confirmed by visual inspection). 
As described in \S~\ref{sect:contslope}, the galaxy spectrum has been adjusted 
to have the same general slope as the model spectrum.
The top left panel of Fig.~\ref{fig:speccompalldoub} shows the spectrum of an 
early-type galaxy, NGC 821 in red, and a 10 Gyr, solar metallicity model 
spectrum overplotted in blue in the region of H$\delta$. The model clearly diverges
 from the data 
redward of H$\delta$ where the red continuum bandpass lies. In the top right 
panel, 
we plot the same two spectra over a larger wavelength region. It is clear that the model and 
galaxy spectra coincide along all other regions of the spectrum, indicating 
that the cause of the divergence redward of H$\delta$ is likely from CN band 
contamination. 

Additionally, we show spectra of M67 and the M31 globular cluster in  the 
middle and bottom panels of Fig.~\ref{fig:speccompalldoub},
respectively, overplotted with corresponding models (again chosen from matching 
Fe4383 and H$\beta$). In Fig.~\ref{fig:wovsfe}, we measured 
consistent ages of M67 regardless of the continuum bandpass used. This result 
was anticipated from the known solar abundance ratios of M67, and so we would 
also expect the corresponding model spectrum and M67 spectrum to be 
nearly indistinguishable. This is clearly the case in the middle panels of  Fig.~\ref{fig:speccompalldoub}. 
On the other hand, the M31 globular cluster diverges from the corresponding 
model redward of H$\delta$. The divergence in the bottom panels 
is even more extreme than in the top panels, presumably from 
the large N abundances in M31 globular clusters found by \citet{Pon98} and \citet{Bur04}. 

\subsection{Additional Index Definitions}
To eliminate the possibility that the above results are specific to the widths and placements
 of the WO97 H$\delta$$_A$ and H$\delta$$_F$ definitions, we have 
defined new indices of varying bandpass widths and placements, as mentioned in 
\S~\ref{sect:definitions}. Although the 
degree of change in derived age differs among these new indices, the overall 
trend remains consistent regardless of the chosen bandpass widths and placements: we derive younger ages when using H$\delta$ indices 
that use the blue shoulder to define the continuum and older ages when using 
the red shoulder. The fact that we were unable to find a satisfactory H$\delta$
 index definition is not suprising, since in \S~\ref{sect:synthres} we found that CN lines 
contaminate H$\delta$ itself, as well as the red continuum side.

\subsection{H$\beta$ Revisited}

We previously demonstrated that while H$\delta$ ages vary substantially, 
depending
on which of the red and/or blue continuum passbands are used, in the
case of H$\beta$ the derived ages are far more self-consistent.  As noted
in \S5.3, however, there is a slight tendency for ages derived from the
integrated spectra of the M31 globular clusters to be older when using the red 
continuum bandpass only and younger when only the blue continuum is used.  While
the effect is subtle, it may signal a small discrepancy between the models and
the M31 metal-rich globular clusters.  As pointed out in \citet{Sch04a},
the Lick H$\beta$ index and its continuum sidebands cover a spectral region in
which a TiO band found in M giants is rapidly changing in depth.  Due to the
great strength of the TiO feature in upper RGB/AGB stars, we suggest that
any inaccuracy in how these advanced stages of stellar evolution 
are
covered in the model isochrones could thus have an impact on the predicted
H$\beta$ index strength.  We also note that in \citet{Sch02} it was
indeed found that the model isochrones do appear to underpredict the
number of RGB stars above the HB level, by a factor of $\sim$2.  

\section{Discussion}

To summarize, we have found that ages derived from the
integrated spectra of early-type galaxies and M31 metal-rich clusters based
on measurements of the H$\delta$ line are highly dependent on how the
continuum is defined from blue and/or red bandpasses.  In contrast, ages
derived from the H$\beta$ feature are robust, regardless of the chosen continuum bandpasses.  We attribute the behavior of 
H$\delta$-derived
ages to the impact of CN molecular lines on the red continuum bandpass and on
H$\delta$ itself.  Specifically, the synthetic spectrum of a CN-enhanced
red giant star, when compared to that of a CN-normal giant, shows that CN
lines are prominent redward of H$\delta$, are present as well in even 
narrowly defined bandpasses centered on H$\delta$ itself, and are substantially weaker in the blue pseudocontinuum of the H$\delta$$_F$ and H$\delta$$_A$ indices.   While CN absorption is 
naturally included in population synthesis models, early-type galaxies and M31 
globular clusters are known to exhibit enhanced CN features resulting from 
non-solar abundance ratios.  It is the non-solar behavior of CN in early-type
galaxies that results in the problematic modelling of H$\delta$.  On the
other hand, we find that H$\delta$ ages are well behaved in both the integrated
spectrum of the open cluster M67 and in the low-mass elliptical galaxy M32,
both of which are known to have primarily solar abundance element ratios.
Thus a consistent picture has emerged in which H$\delta$ ages must be carefully
evaluated when old metal-rich stellar systems with non-solar
abundance ratios are encountered, i.e., massive early-type galaxies at low
redshift.  

Unfortunately, there is no simple way to avoid the CN problem,
since a significant CN molecular line falls within a few \AA \ of the line
center of H$\delta$.  For a typical galaxy velocity dispersion of 200 \kms,
H$\delta$ and the CN line are not resolved.  Thus the only solution is to
eventually model non-solar abundance ratios with sufficient accuracy that the
population synthesis models fully account for the effects of non-solar CN. The difficulties associated with this task were discussed in \S 5.4. Such an effort is beyond the scope of this paper. 
Once such models exist, and  unless non-solar CN is found to be highly correlated with easily
measured elements such as Mg, it will be necessary to independently
measure C and N abundances, which require observations of CH at 4300 \AA \ and
the NH molecular feature at 3360 \AA, a major undertaking. The compensating
 effect of CN lying in both the index itself and the nearby continuum is 
similar in form to the blending of Fe lines with H$\gamma$, so that the 
question arises of whether a method similar to \citet{VA99}'s solution with 
H$\gamma$ can be used in the case of H$\delta$. However, while this is an 
interesting approach for H$\gamma$, defining an index definition for H$\delta$ 
of the same style is severely compromised by the inability to accurately 
     measure both C and N abundances in early-type galaxies.

While CN contamination of H$\delta$ derived ages has been shown to be a major
issue for old stellar populations, it will
have little effect on studies of young populations.   In young populations the
main sequence turnoff is both bright and blue, which means that the integrated
main sequence spectrum is dominated by hot early-type stars that have strong
H$\delta$ features and little or no CN absorption. Moreover, the main sequence
component in young populations
dominates over the contribution from late-type CN-strong RGB stars.  Hence
in young populations the effect of CN on H$\delta$ age determinations is not
an issue.
For example, non-solar abundance ratios should not influence H$\delta$ ages
in studies of k+a galaxies (a common use of the H$\delta$ 
index) due to their relatively young luminosity-weighted ages.   Without 
detailed modelling of non-solar abundance ratio populations, it is impossible
yet to assess at what age the population is young enough to avoid the CN
problem.  However, certainly for populations 2 Gyr and younger, there is
little to worry about.  In the case of M32, with a luminosity-weighted mean
age of 3-4 Gyr, and of M67, with a similar age, we also do not see a problem
with H$\delta$ ages.  However, both of these systems do not exhibit 
substantial non-solar element abundance ratios, thus they do not supply a
critical test.

The difficulty in obtaining reliable ages from H$\delta$ has two major
implications.  First, much interest in H$\delta$ is associated with using it
as the primary age indicator for higher redshift studies, at which H$\beta$
has been redshifted into difficult regions of the night sky emission spectrum.
However, unless the galaxy being studied is at sufficient lookback time that
the luminosity-weighted age is quite small, then CN issues will clearly
hamper the use of H$\delta$ as an age indicator.  Naturally, the emphasis can
shift to H$\gamma$-derived ages, but H$\delta$ can be followed to significantly
higher redshift.  Second, while obtaining a luminosity-weighted mean age for a
galaxy is certainly a valuable first step, in the case when the mean age is
found to be relatively young, the next step is to determine whether
the integrated spectrum is the composite of a young
and an old age population, with uniquely constrained ages.  To do so requires
the use of at least two age-sensitive indices, sufficiently separated in
wavelength that the degeneracy between a single mean age population and two (or more) populations with distinct ages can be overcome. The effect of CN on H$\delta$ ages clearly needs to be taken into consideration if H$\delta$ is to be used for such a subtle distinction.

Finally, the recognition that the H$\delta$ line itself is affected by CN
may well clarify a puzzle regarding the integrated spectra of metal-rich
Galactic globular clusters reported by Rose (1985) and Rose \& Tripicco (1986).
Using line ratio indices that form the ratio in residual central intensities
of neighboring absorption lines, they found that the integrated spectra of
metal-rich Galactic globular clusters form a two-parameter family in their
line ratio index system.  Specifically, clusters with the same value of
the H$\delta$/Fe~I$\lambda$4045 index (which compares the relative central
intensities of H$\delta$ and the Fe~I$\lambda$4045 line) show a spread in
indices that measure CN and Sr~II$\lambda$4077.  Their result is that the
strength of Sr anti-correlates with that of CN.  This is particularly 
puzzling since the ionized strontium line is gravity-sensitive in the same
way as CN.  The result can be understood if the H$\delta$/Fe~I index is 
affected by the same CN overabundance that produces overly strong CN.  The
net effect is to displace the cluster in the diagram of SrII/FeI versus
H$\delta$/FeI, making the cluster appear to be Sr-weak when in fact it is
actually H$\delta$-strong due to the contamination from CN in the H$\delta$
line center.  An implication of this finding is that using Sr~II as a gravity
indicator in the integrated light of early-type galaxies is risky in the
case of those galaxies with non-solar abundances in CN.  Note that this
concern does not apply to the case of M32.

In summary, we warn that when measuring H$\delta$ in composite spectra of
old metal-rich stellar systems with
non-solar abundance ratios, an uncertainty in the derived ages will exist as a result of  unaccounted for CN contamination. 
Thus for low redshift metal-rich
galaxies, H$\beta$ and H$\gamma$ might be preferred as age diagnostics until future work can produce population synthesis models which fully account for non-solar abundancees of CN.
CN abundance issues will not affect H$\delta$ ages for galaxies dominated by
young populations, or for metal-poor systems in which the CN band at 4216 \AA \
is weak.

This study was partially funded by NSF grant AST-0406443
to the University of North Carolina. H.M. acknowledges support from grant HST-GO-104007.01-A.


\input{tab1.tex}


\input{stub.tab2.tex}

\clearpage

\end{document}

%% file: tab1.tex
\pagestyle{empty}

\begin{deluxetable}{lll}
\tabletypesize{\small}
\tablewidth{0pt}
\tablenum{1}
\tablecaption{Index Definitions\label{indexdef}}
\tablehead{
  \colhead{Index}&\colhead{Central Passband}&\colhead{Pseudocontinua}}
\startdata
H$\delta_A$&4041.600-4079.750&4083.500-4122.250, 4128.500-4161.000\tablenotemark{b}\\
H$\delta_{A}b$&4041.600-4079.750&4083.500-4122.250\tablenotemark{b}\\
H$\delta_{A}r$&4041.600-4079.750&4128.500-4161.000\tablenotemark{b}\\
H$\delta_F$&4057.250-4088.500&4091.000-4112.250, 4114.750-4137.250\tablenotemark{b}\\
H$\delta_{F}b$&4057.250-4088.500&4091.000-4112.250\tablenotemark{b}\\
H$\delta_{F}r$&4057.250-4088.500&4114.750-4137.250\tablenotemark{b}\\
H$\beta$&4827.875-4847.875&4847.875-4876.625, 4876.625-4891.625\tablenotemark{a}\\
H$\beta_b$&4827.875-4847.875&4847.875-4876.625\tablenotemark{a}\\
H$\beta_r$&4827.875-4847.875&4876.625-4891.625\tablenotemark{a}\\
Fe4383&4370.375-4221.625&4360.375-4371.625, 4444.125-4456.625\tablenotemark{a}\\
H$\delta_{Nb}$&4093-4106&4083-4091\\
H$\delta_{Wb}$&4091-4113&4040-4088\\
H$\delta_{N}$&4093-4106&4083-4091, 4120-4140\\
H$\delta_{W}$&4091-4113&4040-4088, 4120-4165\\
H$\delta_{Nr}$&4093-4106&4120-4140\\
H$\delta_{Wr}$&4091-4113&4120-4165\\
\enddata
\tablenotetext{a}{Bandpass definitions are from Worthey et al. (1994).}
\tablenotetext{b}{Bandpass definitions are from Worthey \& Ottaviani (1997).}

\label{tab:indexdef}
\end{deluxetable}

%% file: stub.tab2.tex
\pagestyle{empty}

\begin{deluxetable}{lrrrrrrrrrr}
\tabletypesize{\scriptsize}
\tablewidth{0pt}
\tablenum{2}
\tablecolumns{11}
\tablecaption{Index Measurements and Error}
\tablehead{
\colhead{Galaxy ID}& \colhead{H$\delta_Ab$}& \colhead{H$\delta_A$}& \colhead{H$\delta_Ar$}& \colhead{H$\delta_Fb$}& \colhead{H$\delta_F$}& \colhead{H$\delta_Fr$}& \colhead{H$\beta_b$}& \colhead{H$\beta$}& \colhead{H$\beta_r$}& \colhead{Fe4383}}
\startdata
47Tuc & -0.22 & -0.40 & -0.59 & 0.56 & 0.63 & 0.68 & 1.93 & 1.55 & 1.20 & 2.68 \\
\colhead{$\pm$1$\sigma$} & 0.02 & 0.02 & 0.03 & 0.02 & 0.01 & 0.02 & 0.02 & 0.01 & 0.02 & 0.29 \\
M67 & -2.01 & -0.68 & 0.57 & 0.12 & 0.83 & 1.39 & 2.81 & 2.27 & 1.76 & 4.01 \\ 
\colhead{$\pm$1$\sigma$} & 0.15 & 0.14 & 0.17 & 0.10 & 0.08 & 0.09 & 0.11 & 0.09 & 0.12 & 0.19 \\  
\sidehead{M31 Globular Clusters}\\
171-222 & -0.95 & -1.67 & -2.43 & 0.35 & 0.30 & 0.26 & 2.37 & 1.75 & 1.16 & 3.94 \\
\colhead{$\pm$1$\sigma$} & 0.32 & 0.29 & 0.37 & 0.23 & 0.18 & 0.20 & 0.19 & 0.15 & 0.20 & 0.32 \\
163-217 & -1.05 & -2.25 & -3.55 & 0.30 & 0.16 & 0.03 & 2.28 & 1.75 & 1.25 & 4.23 \\
\colhead{$\pm$1$\sigma$} & 0.40 & 0.37 & 0.47 & 0.28 & 0.23 & 0.24 & 0.23 & 0.18 & 0.25 & 0.39 \\
112-174 & -1.50 & -3.26 & -5.24 & 0.06 & -0.31 & -0.63 & 2.19 & 1.50 & 0.84 & 5.03 \\
\colhead{$\pm$1$\sigma$} & 0.44 & 0.40 & 0.51 & 0.30 & 0.25 & 0.26 & 0.23 & 0.18 & 0.25 & 0.48 \\
150-203 & -1.09 & -1.99 & -2.95 & 0.11 & 0.02 & -0.06 & 2.15 & 1.64 & 1.17 & 3.79 \\
\colhead{$\pm$1$\sigma$} & 0.57 & 0.52 & 0.67 & 0.41 & 0.33 & 0.35 & 0.34 & 0.27 & 0.36 & 0.58 \\
193-244 & -0.73 & -2.27 & -3.95 & 0.22 & -0.05 & -0.28 & 2.41 & 1.73 & 1.08 & 4.29 \\
\colhead{$\pm$1$\sigma$} & 0.58 & 0.53 & 0.67 & 0.41 & 0.33 & 0.35 & 0.33 & 0.26 & 0.35 & 0.56 \\
143-198 & -1.31 & -2.43 & -3.64 & 0.17 & 0.05 & -0.04 & 2.27 & 1.59 & 0.95 & 4.44 \\
\colhead{$\pm$1$\sigma$} & 0.35 & 0.32 & 0.41 & 0.25 & 0.20 & 0.21 & 0.20 & 0.16 & 0.22 & 0.34 \\
189-240 & -1.77 & -3.65 & -5.77 & -0.02 & -0.43 & -0.79 & 2.24 & 1.57 & 0.94 & 5.86 \\
\colhead{$\pm$1$\sigma$} & 0.89 & 0.81 & 1.03 & 0.61 & 0.50 & 0.53 & 0.46 & 0.37 & 0.50 & 0.78 \\
\sidehead{CRC03 Early Type Galaxies}\\
A00368p25 & -1.34 & -1.24 & -1.16 & 0.33 & 0.63 & 0.85 & 2.07 & 1.66 & 1.29 & 4.12 \\
\colhead{$\pm$1$\sigma$} & 0.42 & 0.38 & 0.49 & 0.25 & 0.20 & 0.22 & 0.14 & 0.11 & 0.15 & 0.30 \\
A10025 & -2.33 & -1.91 & -1.48 & -0.08 & 0.44 & 0.85 & 1.96 & 1.65 & 1.36 & 4.45 \\
\colhead{$\pm$1$\sigma$} & 0.22 & 0.20 & 0.25 & 0.13 & 0.11 & 0.11 & 0.07 & 0.06 & 0.08 & 0.15 \\
A15572p48 & -1.05 & -0.28 & 0.44 & 0.51 & 0.86 & 1.14 & 2.26 & 1.83 & 1.44 & 3.91 \\
\colhead{$\pm$1$\sigma$} & 0.42 & 0.38 & 0.49 & 0.25 & 0.20 & 0.22 & 0.14 & 0.12 & 0.16 & 0.30 \\
\enddata
\label{tab:galaxyindices}
\end{deluxetable}